# Local cross-sections and energy efficiencies in the multiple optical scattering by two conducting particles

F. G. Mitri

(F.G.Mitri@ieee.org)

**Abstract:** In this work, exact mathematical expansions for the intrinsic electromagnetic (EM) or optical cross-sections (i.e., extinction, scattering and absorption) for a pair of perfectly conducting circular cylinders in a homogeneous medium are derived. The incident illuminating field is an axially–polarized electric field composed of plane travelling waves with an arbitrary angle of incidence in the polar plane. The formalism is based on the multipole modal expansion method in cylindrical coordinates and the translational addition theorem applicable to any range of frequencies. An effective EM field, incident on the probed cylinder, is defined first, which includes the initial and re-scattered field from the second cylinder. Subsequently, it is used jointly with the scattered field to derive the mathematical expressions for the intrinsic/local cross–sections based on integrating their corresponding time-averaged intensities over the surface of the probed object by applying the Poynting theorem. Numerical computations for the intrinsic extinction (or scattering) energy efficiencies for a pair of conducting circular cylinders with different radii in a homogeneous medium are considered. Emphasis is given on varying the interparticle distance, the angle of incidence, and the dimensionless sizes of the cylinders. The results computed *a priori* can be useful in the full characterization of a multiple scattering system of many particles, in conjunction with experimental data for the extrinsic cross-sections.

## 1. Introduction

The absorption, extinction and scattering cross-sections [1] (or energy efficiencies) are physical observables used for quantitative characterization in the interaction of electromagnetic (EM) [or light] waves with an object [2-6] or multiple objects. For example, they can be used in the accurate predictions of the interaction properties of aerosol particles in a cluster [7].

Acknowledging the advantages of developing rigorous analytical models to predict and compute such physical observables in multiple scattering systems by several objects [7-21], various investigations [22-26] considered the modal expansion method with boundary matching in cylindrical coordinates for this purpose. In these works, analytical/mathematical equations for the *extrinsic* cross-sections have been developed, which provide quantifiable information on the global characteristics of the multiple interacting system of particles with the incident field. For example, the obtained extrinsic scattering cross-section accounts for the multiple/cross-interferences, with a clear dependence on the distance between two interacting objects. It can be correlated with experimental measures in the far-field. This has been acknowledged recently in the multiple acoustical scatterings by rigid and fluid cylinders [27, 28]. This approach, however, is not adequate to quantify the *inherent/intrinsic* scattering, absorption and extinction properties, which exhibit local properties for each particle in the presence of other particles, quite distinguishable from the extrinsic ones. Thus, it is of some importance to develop a new analytical model for the intrinsic extinction, scattering and absorption cross-sections in order to predict the local properties of the multiple interacting system of particles with the incident EM/optical field.

The aim of this investigation is therefore directed toward this goal, where exact partial-wave series expansions for the intrinsic scattering, absorption and extinction cross-sections in the multiple scattering by a pair of perfectly conducting circular cylinders of different radii are presented. In this procedure, an *effective* EM field incident on the probed cylinder includes the initial and re-scattered field from the second cylinder. It is used subsequently with the field scattered from the probed object itself to obtain the series expansions for the local/inherent cross-sections. Using the modal expansion method with boundary matching in cylindrical coordinates in conjunction with the translational addition theorem, mathematical expressions involving the interparticle distance, the angle of incidence as well as the radii of both cylinders are established. Numerical computations illustrate the analysis for two conducting circular cylinders in a homogeneous non-magnetic free-space.

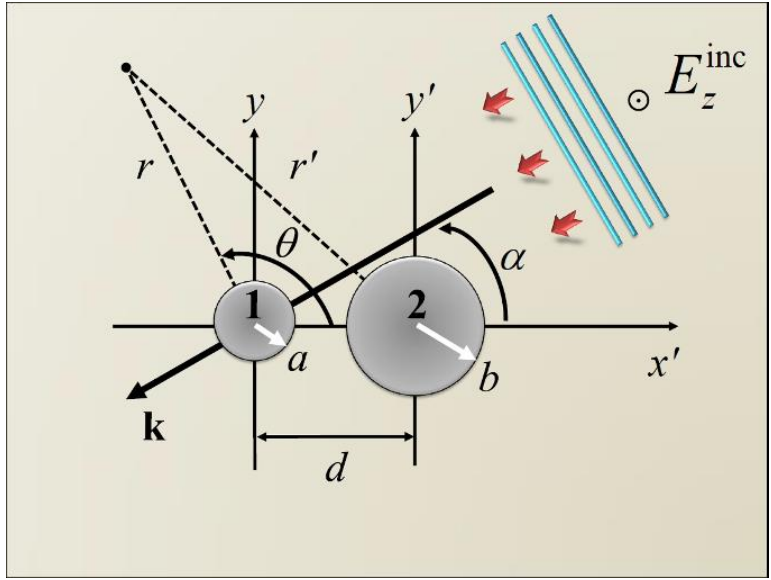

**Fig. 1.** Graphical sketch representing the interaction of an axially-polarized (i.e. along the *z*-direction) electric field of electromagnetic monochromatic plane waves of arbitrary incidence angle $\alpha$ in the polar plane with a pair of conducting cylinders of different radii *a* and *b*, respectively, in a homogeneous unbounded space. The parameter *d* is the interparticle distance between the center of mass of both cylinders.

The paper is organized as follows: section 2 provides a comprehensive description of the modal expansion method and the essential mathematical recipe to obtain the exact mathematical expressions for the extinction, absorption and scattering cross-sections. Section 3 presents and discusses the results of the numerical computations for the related energy efficiencies, obtained from the expressions of the cross-sections. Finally, section 4 summarizes the findings of this investigation.

## 2. Method

The procedure of determining the expressions for the intrinsic scattering, absorption and extinction cross-sections consists first of obtaining the expressions for the effective incident and scattered fields for the probed cylinder [29]. Therefore, an analysis for the multiple scattering of the incident field by the pair of conducting cylinder of different radii (equivalent to its acoustic counterpart [30]) is required.

Consider an electric field with axial polarization (i.e. along the *z*-direction – see Fig. 1) composed of plane progressive waves. The field is incident upon a pair of two perfectly conducting circular cylinders of different radii *a* and *b*, respectively. The interparticle distance *d* separates the center of mass of the cylinders, and the field has an angle of incidence $\alpha$ in the polar plane as shown in Fig. 1. The waves propagate in a non-conducting nonmagnetic homogenous medium without any dissipation (i.e. lossless medium). The axial component for the incident electric field is expressed as

$$E_z^{\text{inc}} = E_0 e^{-i\omega t} e^{i\mathbf{k}\cdot\mathbf{r}}, \qquad (1)$$

where $E_0$ is the field amplitude, $\omega$ is the angular frequency, **r** is the vector position, and **k** is the wave-vector.

In the polar system $(r,\theta)$ centered on the first circular cylinder, Eq.(1) is expanded as a partial-wave series as

$$E_z^{\text{inc}}(r,\theta,t) = E_0 e^{-i\omega t} e^{-ikr\cos(\theta-\alpha)}$$

$$= E_0 e^{-i\omega t} \sum_{n=-\infty}^{+\infty} i^{-n} e^{-in\alpha} J_n(kr) e^{in\theta}, \quad (2)$$

where $k = \sqrt{\varepsilon}\,\omega/c$ is the wavenumber, $c$ is the speed of light in the medium of wave propagation, and $J_n(\cdot)$ is the cylindrical Bessel function of the first kind.

Based on Faraday's law, the corresponding incident polar component of the magnetic field is obtained as [31-33],

$$H_\theta^{\text{inc}}(r,\theta,t) = i\sqrt{\varepsilon} E_0 e^{-i\omega t} \sum_{n=-\infty}^{+\infty} i^{-n} e^{-in\alpha} J_n'(kr) e^{in\theta}, \quad (3)$$

where $\varepsilon$ is the dielectric constant of the medium of wave propagation. The prime in Eq.(3) denotes a derivative with respect to the argument of the cylindrical Bessel function.

The scattered fields from both cylinders are also expressed in partial-wave series expansions. The axial scattered electric field component $E_{z,1}^{\text{sca}}$ from the first cylinder in the system of coordinates $(r, \theta)$, and the second, $E_{z,2}^{\text{sca}}$ from the second cylinder in the system of coordinates $(r', \theta')$ are expressed, respectively, as

$$E_{z,1}^{\text{sca}}(r,\theta,t) = E_0 e^{-i\omega t} \sum_{n=-\infty}^{+\infty} C_{n,1} H_n^{(1)}(kr) e^{in\theta}, \quad (4)$$

$$E_{z,2}^{\text{sca}}(r',\theta',t) = E_0 e^{-i\omega t} \sum_{n=-\infty}^{+\infty} C_{n,2} H_n^{(1)}(kr') e^{in\theta'}, \quad (5)$$

where $C_{n,1}$ and $C_{n,2}$ are the modal coefficients, which depend on the material properties of the cylinders, and are coupled with the incident field. $H_n^{(1)}(.)$ is the cylindrical Hankel function of the first kind of order $n$.

Eq.(5) shows that the scattered axial electric field component from the second cylinder is written in terms of the primed system of coordinates. To determine the coupled expansion coefficients, suitable boundary matching conditions should be satisfied, all written in the same system of coordinates. Therefore, the translational addition theorem is utilized for this purpose. Accordingly, the following equalities hold [34],

$$H_n^{(1)}(kr')e^{in\theta'} = \begin{cases} \sum_{m=-\infty}^{+\infty} J_{m-n}(kd) H_m^{(1)}(kr) e^{im\theta}, & r > d \\ \sum_{m=-\infty}^{+\infty} J_m(kr) H_{m-n}^{(1)}(kd) e^{im\theta}, & r < d \end{cases}, \quad (6)$$

$$H_n^{(1)}(kr)e^{in\theta} = \begin{cases} \sum_{m=-\infty}^{+\infty} J_{n-m}(kd) H_m^{(1)}(kr') e^{im\theta'}, & r' > d \\ \sum_{m=-\infty}^{+\infty} J_m(kr') H_{n-m}^{(1)}(kd) e^{im\theta'}, & r' < d \end{cases}. \quad (7)$$

It follows that the total axial electric field component in the host medium of wave motion close to the first (second, respectively) cylinder such that $r < d$ ($r' < d$, respectively) is rewritten in either coordinate system as,

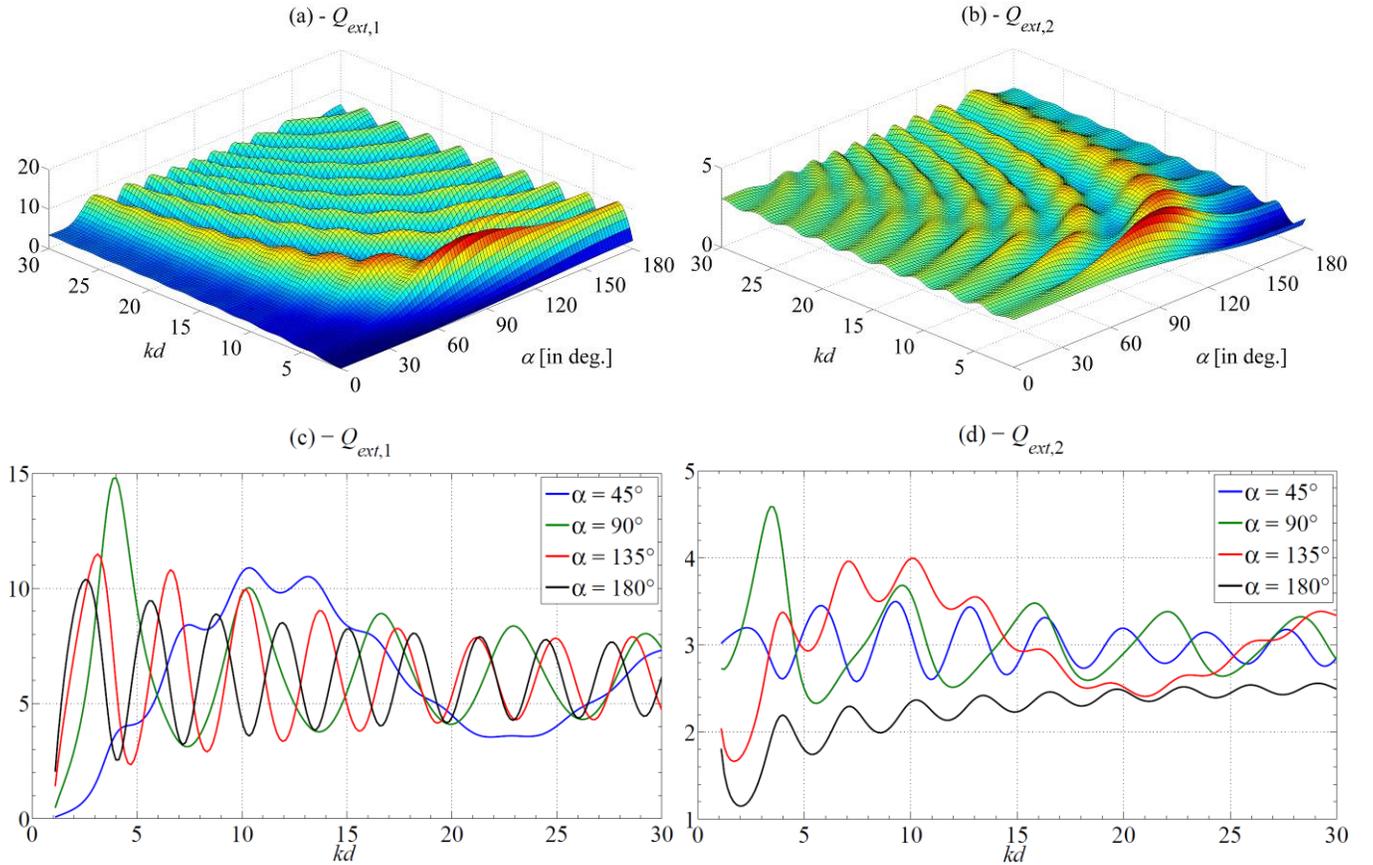

**Fig. 2.** Panels (a) and (b) display the plots for the intrinsic extinction energy efficiencies for two perfectly conducting circular cylinders having $ka = 0.1$ and $kb = 1$ versus the dimensionless interparticle distance $kd$ and incidence angle $\alpha$.

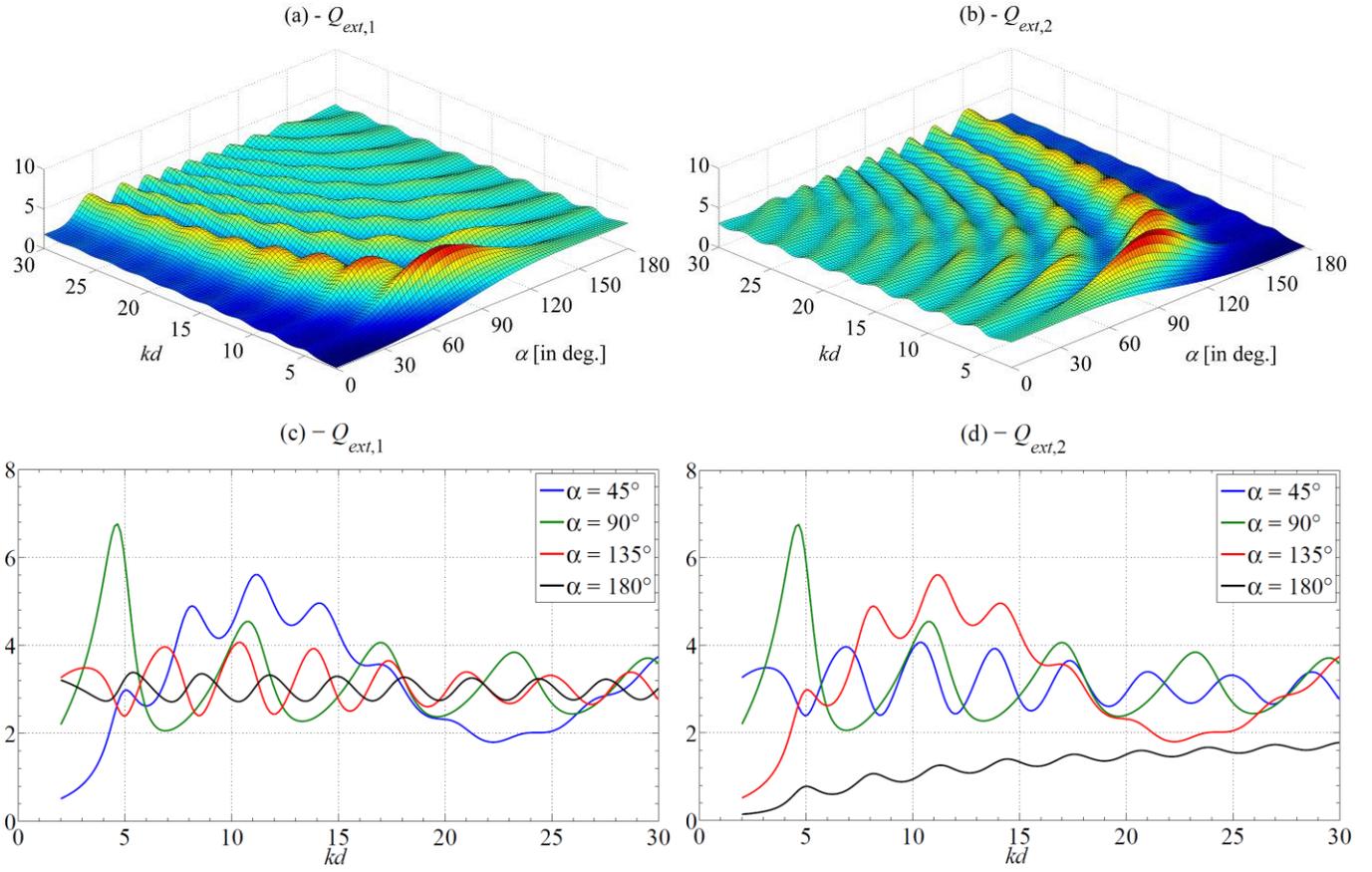

**Fig. 3.** The same as in Fig. 2, but $ka = kb = 1$.

$$E_z^{\text{tot}}(r,\theta,t)\big|_{r<d} = E_z^{\text{inc}}(r,\theta,t) + E_{z,1}^{\text{sca}}(r,\theta,t) + E_{z,2}^{\text{sca}}(r,\theta,t)$$

$$= E_0 e^{-i\omega t}\left[\sum_{n=-\infty}^{+\infty}\left(i^{-n}e^{-in\alpha}J_n(kr) + C_{n,1}H_n^{(1)}(kr)\right)e^{in\theta}\right. \quad (8)$$

$$\left. + \sum_{n=-\infty}^{+\infty}\left(C_{n,2}\sum_{m=-\infty}^{+\infty}J_m(kr)H_{m-n}^{(1)}(kd)e^{im\theta}\right)\right],$$

$$E_z^{\text{tot}}(r',\theta',t)\big|_{r'<d} = E_z^{\text{inc}}(r',\theta',t) + E_{z,1}^{\text{sca}}(r',\theta',t) + E_{z,2}^{\text{sca}}(r',\theta',t)$$

$$= E_0 e^{-i\omega t}\left[\sum_{n=-\infty}^{+\infty}\left(i^{-n}e^{-in\alpha}e^{-ikd\cos\alpha}J_n(kr') + C_{n,2}H_n^{(1)}(kr')\right)e^{in\theta'}\right. \quad (9)$$

$$\left. + \sum_{n=-\infty}^{+\infty}\left(C_{n,1}\sum_{m=-\infty}^{+\infty}J_m(kr')H_{n-m}^{(1)}(kd)e^{im\theta'}\right)\right].$$

Since the incident electric field is axially polarized (perpendicularly to the polar plane – known also as TM polarization), the boundary conditions that must be satisfied on the surface of each perfectly conducting cylinder require that [30]

$$E_z^{tot}(r,\theta,t)\big|_{r=a} = 0, \quad (10)$$

$$E_z^{tot}(r',\theta',t)\big|_{r'=b} = 0, \quad (11)$$

where *a* and *b* are the radii of cylinders 1 and 2, respectively (Fig. 1).

Using the orthogonality condition, the coupled linear system of equations is obtained from Eqs.(10) and (11) as follows,

$$i^{-\ell}e^{-i\ell\alpha}J_\ell(ka) + C_{\ell,1}H_\ell^{(1)}(ka) + J_\ell(ka)\sum_{n=-\infty}^{\infty} C_{n,2}H_{\ell-n}^{(1)}(kd) = 0, \quad (12)$$

$$i^{-\ell}e^{-i\ell\alpha}e^{-ikd\cos\alpha}J_\ell(kb) + C_{\ell,2}H_\ell^{(1)}(kb) + J_\ell(kb)\sum_{n=-\infty}^{\infty} C_{n,1}H_{n-\ell}^{(1)}(kd) = 0. \quad (13)$$

This system of equations is solved by matrix inversion procedures such that the modal coefficients $C_{n,1}$ and $C_{n,2}$ can be determined numerically by imposing a truncation limit in the series such that $N_{max}$ = [max(max(*ka*, *kb*), *kd*)] + 30. This limit ensures suitable convergence of the series and negligible truncation numerical error [30].

As discussed previously, the modal coefficients depend on *both* objects as shown from Eqs.(12) and (13). Column vectors are of dimension 1 × (2$N_{max}$ + 1) elements, while the square matrices are of dimension (2$N_{max}$ + 1) × (2$N_{max}$ + 1) elements.

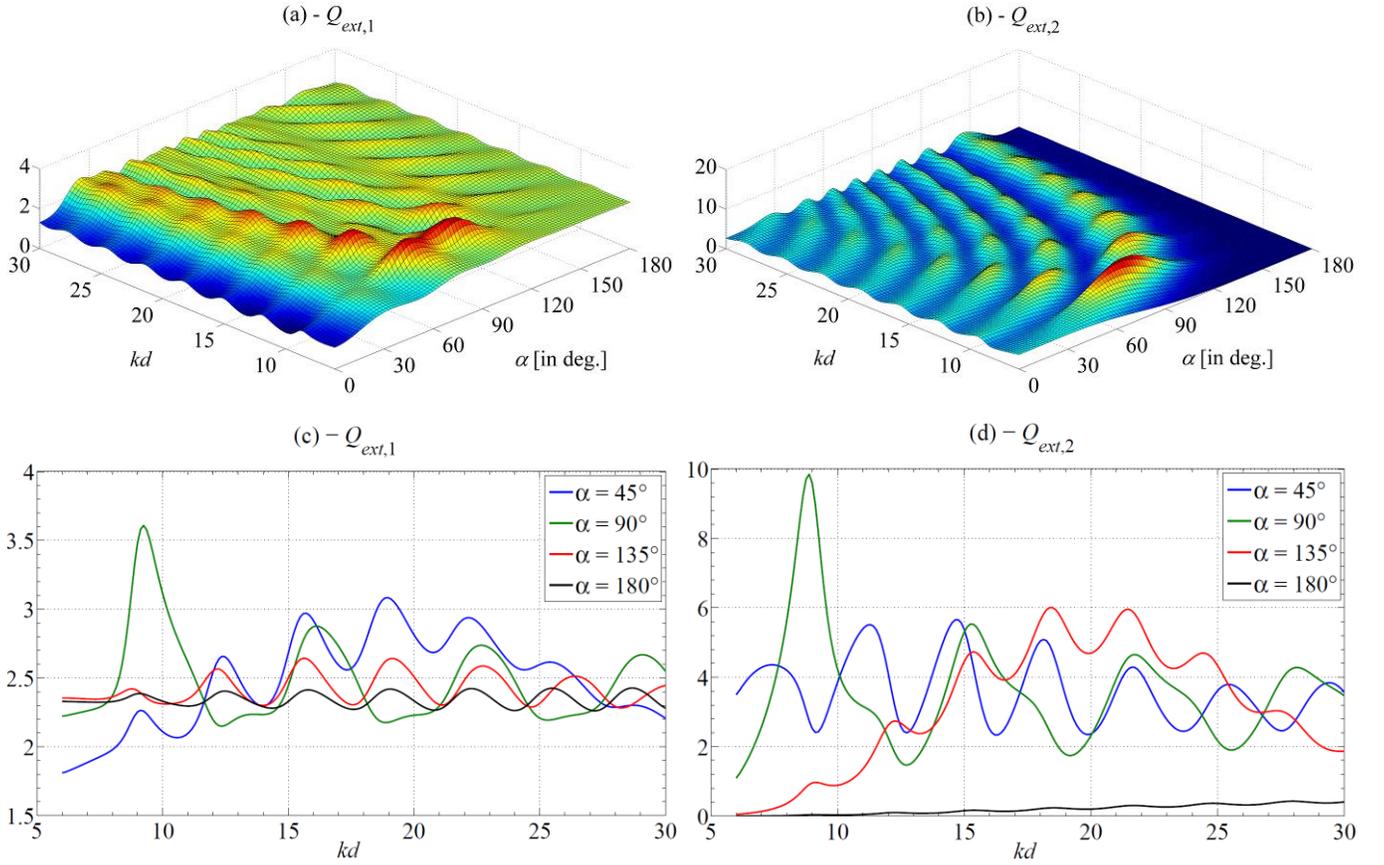

**Fig. 4.** The same as in Fig. 2, but $ka = 5$ and $kb = 1$.

The procedure of determining the expressions for the intrinsic extinction, scattering and absorption cross-sections consists of integrating the associated time-averaged intensities over the surface of each cylinder. When multiple scattering effects arise between the particle pair, the probed cylinder will be subjected to an incident *effective* field composed of the primary incident EM field in addition to the multiple re-scattering field due to the presence of the other cylinder. Therefore, to obtain the mathematical expressions for the cross-sections related to a particular cylinder, an *effective* incident field on a particular scatterer should be defined and determined initially. This method provides the adequate means to obtain the intrinsic cross-sections on the probed cylinder as the problem in hand is reduced to that of the single object case, but with an incident *effective* field.

The determination of the incident effective electric (and magnetic) fields on each cylinder in each system of coordinates is accomplished based on Eqs.(8) and (9) by recognizing that the scattered (outgoing) waves from the probed particle cannot be simultaneously directed toward its own surface. Therefore, the effective electric field component due to the primary incident waves in the vicinity of the first cylinder (i.e. $r < d$) in the presence of the second cylinder is obtained from Eq.(8) as,

$$E_{z,1,\text{eff}}^{\text{inc}}(r,\theta,t) = E_0 e^{-i\omega t}\left[\sum_{n=-\infty}^{+\infty} i^{-n} e^{-in\alpha} J_n(kr) e^{in\theta} \right.$$
$$\left. + \sum_{n=-\infty}^{+\infty}\left(C_{n,2} \sum_{m=-\infty}^{+\infty} J_m(kr) H_{m-n}^{(1)}(kd) e^{im\theta}\right)\right]. \qquad (14)$$

In the primed system of coordinates, the effective electric field component incident on the second cylinder is deduced from Eq.(9) such that,

$$E_{z,2,\text{eff}}^{\text{inc}}(r',\theta',t) = E_0 e^{-i\omega t}\left[\sum_{n=-\infty}^{+\infty} i^{-n} e^{-in\alpha} e^{-ikd\cos\alpha} J_n(kr') e^{in\theta'} \right.$$
$$\left. + \sum_{n=-\infty}^{+\infty}\left(C_{n,1}\sum_{m=-\infty}^{+\infty} J_m(kr') H_{n-m}^{(1)}(kd) e^{im\theta'}\right)\right]. \quad (15)$$

Subsequently, the formulation of the scattering in the far-field [35] can be used, since the problem in hand is now reduced to the single-object case. This procedure leads to exact mathematical expressions without any approximations.

In the far-field limit, the cylindrical Bessel and Hankel functions of the first kind take the following forms, respectively, $J_n(kr) \underset{kr\to\infty}{\approx} \sqrt{\frac{2}{\pi kr}}\cos\left(kr - \frac{n\pi}{2} - \frac{\pi}{4}\right)$ and $H_n^{(1)}(kr) \underset{kr\to\infty}{\approx} \sqrt{\frac{2}{\pi kr}} e^{i(kr - \frac{n\pi}{2} - \frac{\pi}{4})}$.

The expression for the intrinsic scattering cross-sections for the cylinders 1 and 2 can be obtained from the previous analysis [35] for the single scatterer in arbitrary-shaped beams as

$$\sigma_{sca,\{1,2\}} \underset{(kr,kr')\to\infty}{=} -\frac{1}{2I_0}\iint_S \Re\left\{E_{z,\{1,2\}}^{\text{sca}} H_{\theta,\{1,2\}}^{\text{sca}*}\right\} dS, \quad (16)$$

where $H_{\theta,\{1,2\}}^{\text{sca}}$ is obtained from Faraday's law [as in Eq.(3)], the superscript * denotes a complex conjugate, and $I_0 = \sqrt{\varepsilon}|E_0|^2/2$ is the characteristic intensity.

The expression for the intrinsic absorption cross-section is also determined from [35] by noticing that the polarization of the incident plane waves is TM (i.e., axial polarization of the incident field). Therefore, based on Eqs.(9)-(11) in [35], the intrinsic absorption cross-section is obtained as

$$\sigma_{abs,\{1,2\}} \underset{(kr,kr')\to\infty}{=} \frac{1}{2I_0}\iint_S \Re\left\{\begin{array}{l} E_{z,\{1,2\}}^{\text{sca}} H_{\theta,\{1,2\}}^{\text{sca}*} \\ + E_{z,\{1,2\},\text{eff}}^{\text{inc}} H_{\theta,\{1,2\}}^{\text{sca}*} \\ + E_{z,\{1,2\}}^{\text{sca}} H_{\theta,\{1,2\},\text{eff}}^{\text{inc}*} \end{array}\right\} dS, \quad (17)$$

where $H_{\theta,\{1,2\},\text{eff}}^{\text{inc}}$ is also obtained based on Faraday's law.

Finally, the expression for the extinction cross-section is obtained from Eqs.(16) and (17) as

$$\sigma_{ext,\{1,2\}} \underset{(kr,kr')\to\infty}{=} \sigma_{sca,\{1,2\}} + \sigma_{abs,\{1,2\}}$$
$$\underset{(kr,kr')\to\infty}{=} \frac{1}{2I_0}\iint_S \Re\left\{E_{z,\{1,2\},\text{eff}}^{\text{inc}} H_{\theta,\{1,2\}}^{\text{sca}*} + E_{z,\{1,2\}}^{\text{sca}} H_{\theta,\{1,2\},\text{eff}}^{\text{inc}*}\right\} dS. \quad (18)$$

Substituting the expressions for the electric and magnetic field components into Eqs.(16)-(18) and performing the adequate algebraic manipulation using the property of the angular integral $\int_0^{2\pi} e^{i(n-\ell)\theta} d\theta = 2\pi\delta_{n,\ell}$, where $\delta_{ij}$ is the Kronecker delta function and $dS = rd\theta$ (in 2D), the intrinsic cross-sections (per unit-length) are expressed as exact partial-wave series as,

$$\sigma_{sca,1} = \frac{4}{k}\sum_{n=-\infty}^{\infty}|C_{n,1}|^2, \quad (19)$$

$$\sigma_{abs,1} = -\frac{4}{k}\Re\left\{\sum_{n=-\infty}^{\infty} C_{n,1}^*\left[\mathrm{i}^{-n}\mathrm{e}^{-in\alpha} + \sum_{m=-\infty}^{\infty} C_{m,2} H_{n-m}^{(1)}(kd)\right] + |C_{n,1}|^2\right\}, \qquad (20)$$

$$\sigma_{ext,1} = -\frac{4}{k}\Re\left\{\sum_{n=-\infty}^{\infty} C_{n,1}^*\left[\mathrm{i}^{-n}\mathrm{e}^{-in\alpha} + \sum_{m=-\infty}^{\infty} C_{m,2} H_{n-m}^{(1)}(kd)\right]\right\}, \qquad (21)$$

$$\sigma_{sca,2} = \frac{4}{k}\sum_{n=-\infty}^{\infty} |C_{n,2}|^2, \qquad (22)$$

$$\sigma_{abs,2} = -\frac{4}{k}\Re\left\{\sum_{n=-\infty}^{\infty} C_{n,2}^*\left[\begin{array}{l}\mathrm{i}^{-n}\mathrm{e}^{-in\alpha}\mathrm{e}^{-ikd\cos\alpha} \\ + \sum_{m=-\infty}^{\infty} C_{m,1} H_{m-n}^{(1)}(kd)\end{array}\right] + |C_{n,2}|^2\right\}, \qquad (23)$$

$$\sigma_{ext,2} = -\frac{4}{k}\Re\left\{\sum_{n=-\infty}^{\infty} C_{n,2}^*\left[\mathrm{i}^{-n}\mathrm{e}^{-in\alpha}\mathrm{e}^{-ikd\cos\alpha} + \sum_{m=-\infty}^{\infty} C_{m,1} H_{m-n}^{(1)}(kd)\right]\right\}. \qquad (24)$$

Subsequently, the corresponding *dimensionless* energy efficiency factors are determined from Eqs.(19)-(24), such that,

$$Q_{\{sca,abs,ext\},1} = \sigma_{\{sca,abs,ext\},1}/(2a), \qquad (25)$$

$$Q_{\{sca,abs,ext\},2} = \sigma_{\{sca,abs,ext\},2}/(2b). \qquad (26)$$

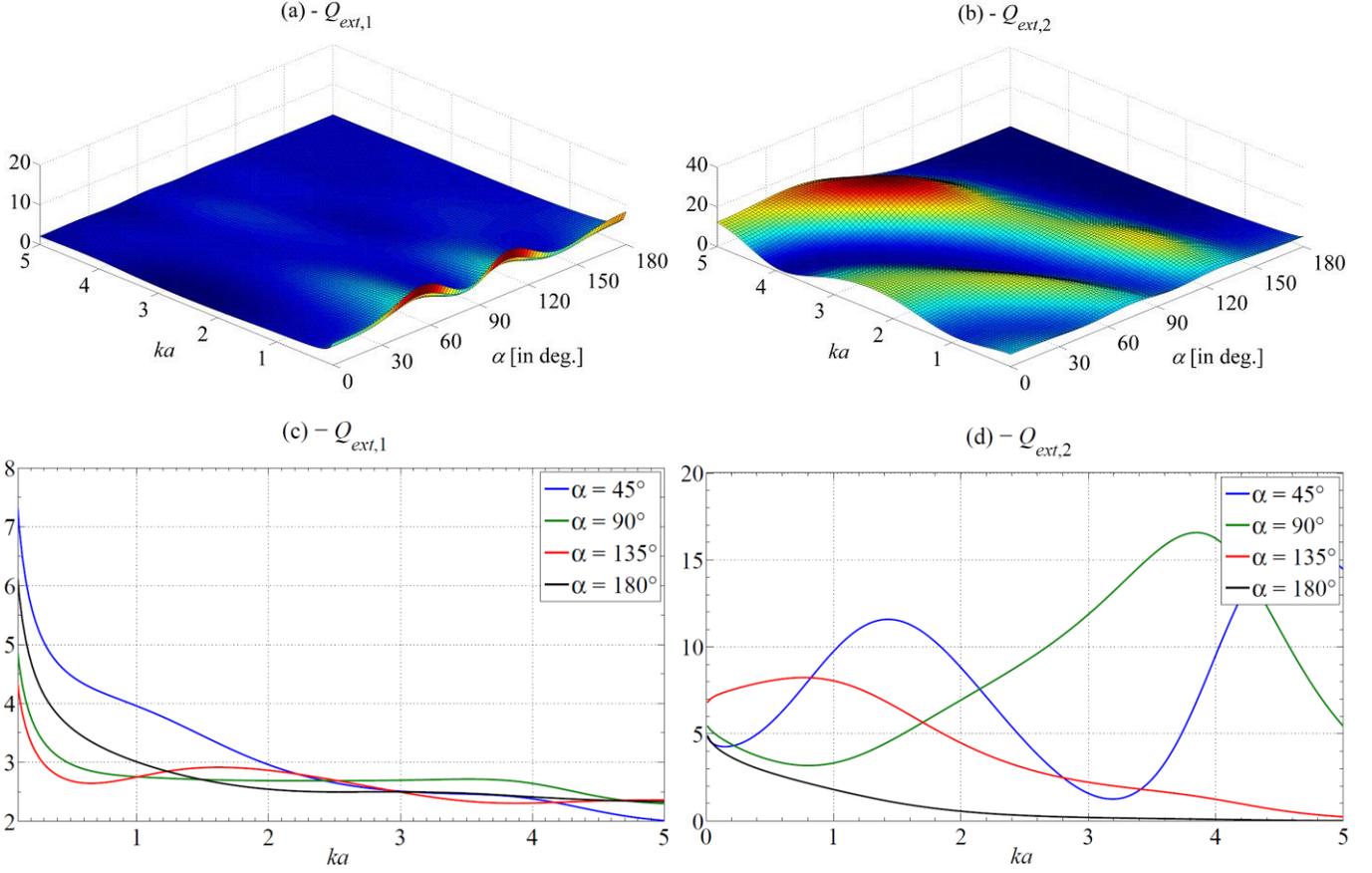

**Fig. 5.** Panels (a) and (b) show the plots for the local extinction energy efficiencies for a pair of perfectly conducting circular cylinders versus the dimensionless size parameter $ka$ and incidence angle $\alpha$, for $kb = 0.1$ and $kd = 7$.

## 3. Numerical examples

In the following, numerical examples for the intrinsic energy efficiencies given by Eqs.(25) and (26) illustrate the analysis for a pair of conducting cylinders of radii $a$ and $b$ respectively, where the separation distance between their center of mass is denoted by $d$ (Fig. 1). Since the cylinders are considered perfectly conducting (and the medium of wave propagation is vacuum, i.e. $\varepsilon = 1$), the following properties hold, that is, $\sigma_{abs,\{1,2\}} = 0$, and $\sigma_{ext,\{1,2\}} = \sigma_{sca,\{1,2\}}$. Therefore, only the intrinsic extinction energy efficiency for each cylinder is computed.

The examples chosen subsequently consider a Rayleigh (small) cylinder, cylinders of equal radii, and Mie (large) cylinders. The plots considered variations of the dimensionless interparticle distance $kd$ and angle of incidence for various size parameters of one of the cylinders. Moreover, additional plots considered the variations of the intrinsic extinction/scattering efficiency at various interparticle distances. Those examples illustrate the analysis for different scenarios that may be considered in practice.

Panels (a) and (b) of Fig. 2 show the results for the intrinsic extinction (or scattering) efficiencies for both cylinders, such that $ka = 0.1$ and $kb = 1$, respectively, in the ranges $0 \leq \alpha \leq 180°$ and $(ka + kb) < kd \leq 30$ for an axially-polarized electric field component of plane progressive waves. Panels (c) and (d) display the one-line profile plots versus $kd$ for four different incidence angles, $\alpha = 45°$, $90°$, $135°$ and $180°$. At $\alpha = 0°$ in panel (a), the first Rayleigh (i.e., small) cylinder with $ka = 0.1$ is in the shadow area of the other cylinder with $kb = 1$, such that the intrinsic extinction $Q_{ext,1}$ by the first cylinder is small in amplitude. As $\alpha$ increases > 30°, larger amplitudes are manifested in the plot. Moreover, periodic oscillations are exhibited as both $\alpha$ and $kd$ increase. At $\alpha = 180°$, the incident field interacts first with the small cylinder, leading to a larger amplitude variation of $Q_{ext,1}$. In panel (b), a different behavior is anticipated such that the emergence of amplitude variations occurs at $\alpha = 0°$ and weakens when $\alpha$ reaches 180°. In both panels, the amplitude variations versus $\alpha$ and $kd$ are related to the strength of the many interactions between the two conducting cylinders. The waves scattered from the first impenetrable perfectly conducting cylinder induce reflected waves (i.e., specular reflection) from the second cylinder along with circumferential circumnavigating waves (i.e., Franz waves [36]) propagating in the exterior homogeneous medium. This phenomenon occurs repetitively during the multiple scattering process. It enhances or reduces the intrinsic extinction (or scattering) energy efficiency factor. From geometry considerations as shown in Fig. 1, clearly the angle of incidence modifies the path length for the circumnavigating Franz waves and their interactions/interferences with the specularly reflected waves, as $\alpha$ varies.

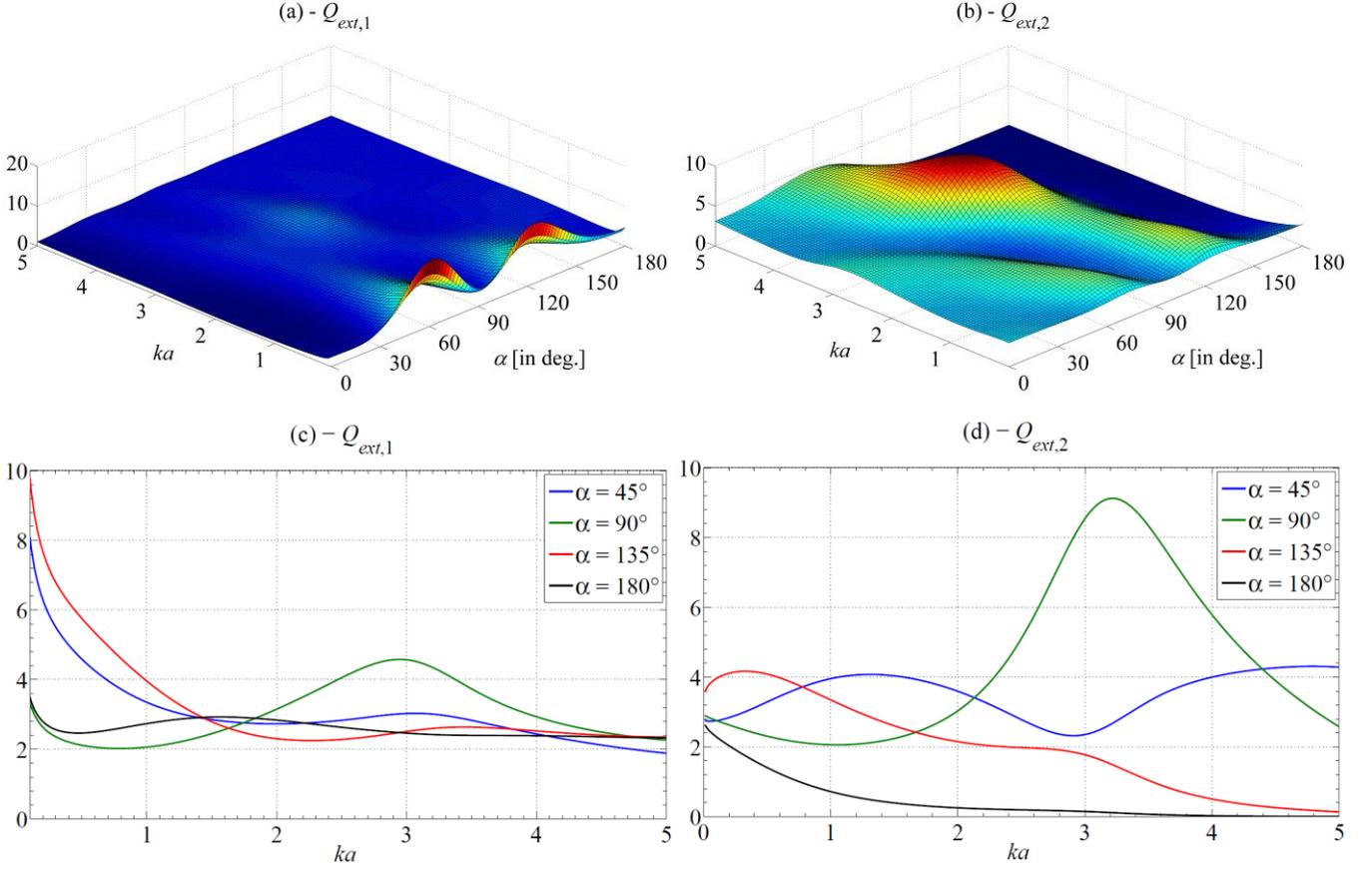

**Fig. 6.** The same as in Fig. 5 but $kb = 1$.

The increase of the size parameter $ka$ for the first particle is also considered such that $ka = kb = 1$. The related results are exhibited in panels (a) and (b) of Fig. 3. Moreover, panels (c) and (d) display the one-line profile plots versus $kd$ for four different incidence angles, $\alpha$ = 45°, 90°, 135° and 180°. The comparison between the plots in each panel demonstrates that for this particular case where $ka = kb$, the following property for the local extinction energy efficiency factors holds, such that, $Q_{ext,1}(\alpha, kd)\big|_{ka=kb} = Q_{ext,2}(\pi - \alpha, kd)\big|_{ka=kb}$. Thus, at $\alpha$ = 90°, $Q_{ext,1} = Q_{ext,2}$ when $ka = kb$.

The non-dimensional size of the first cylinder is further increased to $ka = 5$, whereas $kb = 1$. The related computational plots for the local extinction (or scattering) energy efficiencies are displayed in panels (a) and (b) of Fig. 4, respectively, while panels (c) and (d) display the one-line profile plots versus $kd$ for four different incidence angles, $\alpha$ = 45°, 90°, 135° and 180°. In this case, the multiple scatterings exhibited by the number of the amplitude variations in the plots are changed as $ka$ (or $kb$) varies compared to the previous plots in Figs. 2 and 3.

Consider now the case of a small Rayleigh conducting cylinder such that $kb = 0.1$ at a fixed dimensionless interparticle distance $kd = 7$ while varying the size parameter $ka$ of the first cylinder as well as the angle of incidence. Panels (a) and (b) of Fig. 5 display the plots based on the computations of Eqs.(25) and (26), in the ranges $0° \leq \alpha \leq 180°$ and $0 < ka \leq 5$, while panels (c) and (d) display the one-line profile plots versus $ka$ for four different incidence angles, $\alpha$ = 45°, 90°, 135° and 180°. As shown in panel (a), $Q_{ext,1}$ exhibits the large amplitude variations at low $ka$ values only, while $Q_{ext,2}$ is significantly affected for incidence angles $0 \leq \alpha \leq 150°$.

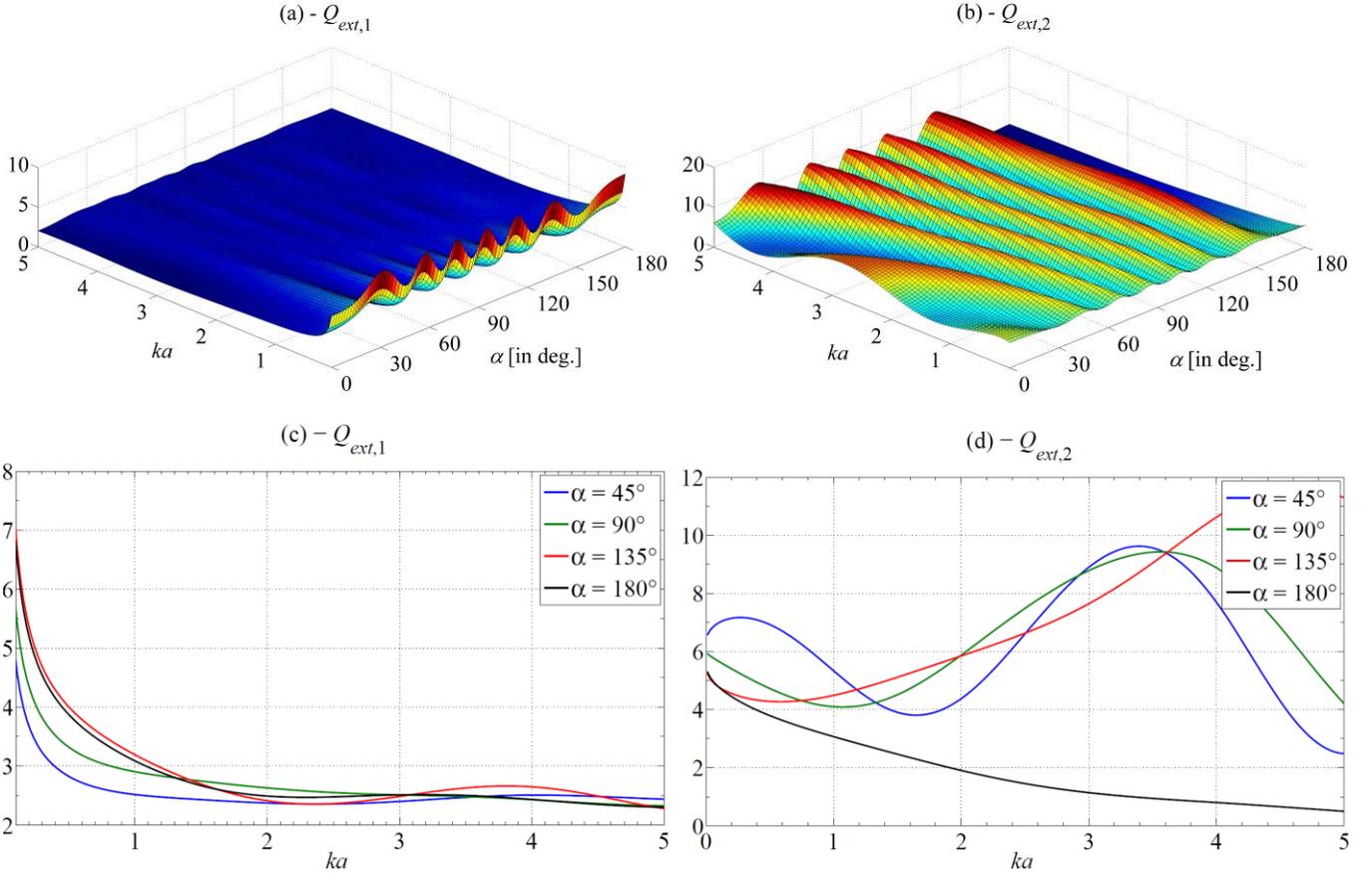

**Fig. 7.** The same as in Fig. 5, but $kd = 20$.

The effect of increasing $kb$ (= 1) is shown in the plots of panels (a)-(d) of Fig. 6, which display similar behaviors to those shown previously in Fig. 5. Nonetheless, the amplitude variations in the plots are different.

The effect of increasing the dimensionless interparticle distance to $kd = 20$ is further analyzed for the second Rayleigh (small) cylinder ($kb = 0.1$, Fig. 7) and for a larger one ($kb = 1$, Fig. 8). As noticed from the plots in Figs. 7 and 8, the interparticle distance, the angle of incidence as well as the size parameters of both cylinders alter the behavior of the intrinsic extinction (or scattering) energy efficiency factors, where multiple amplitude variations arise. As $kd$ increases, the variations in the amplitudes become more frequent. Although $kb$ is fixed while computing $Q_{ext,2}$ in Figs. 5-8, it still depends on $ka$ as shown in the coupled system of equations (12) and (13).

## 4. Conclusion and perspectives

In summary, exact mathematical series expansions are derived for the intrinsic/local extinction, absorption and scattering cross-sections and their related energy efficiency factors. The modal multipole expansion method in cylindrical coordinates is used in conjunction with the translational addition theorem. An effective incident field is determined first, then used to obtain the series expansions for the intrinsic/local cross-sections based on Poynting's theorem. Assuming an axially-polarized electric field component composed of plane travelling waves of arbitrary incidence angle in the polar plane, numerical computations for a pair of perfectly conducting cylinders in a homogeneous free-space are presented and discussed. Particular emphases are given on the interparticle distance separating both cylinders, the angle of incidence as well as their sizes. The present formalism sheds light on the intrinsic (i.e., local) extinction or scattering properties in a system containing a pair of interacting cylindrical particles in EM waves.

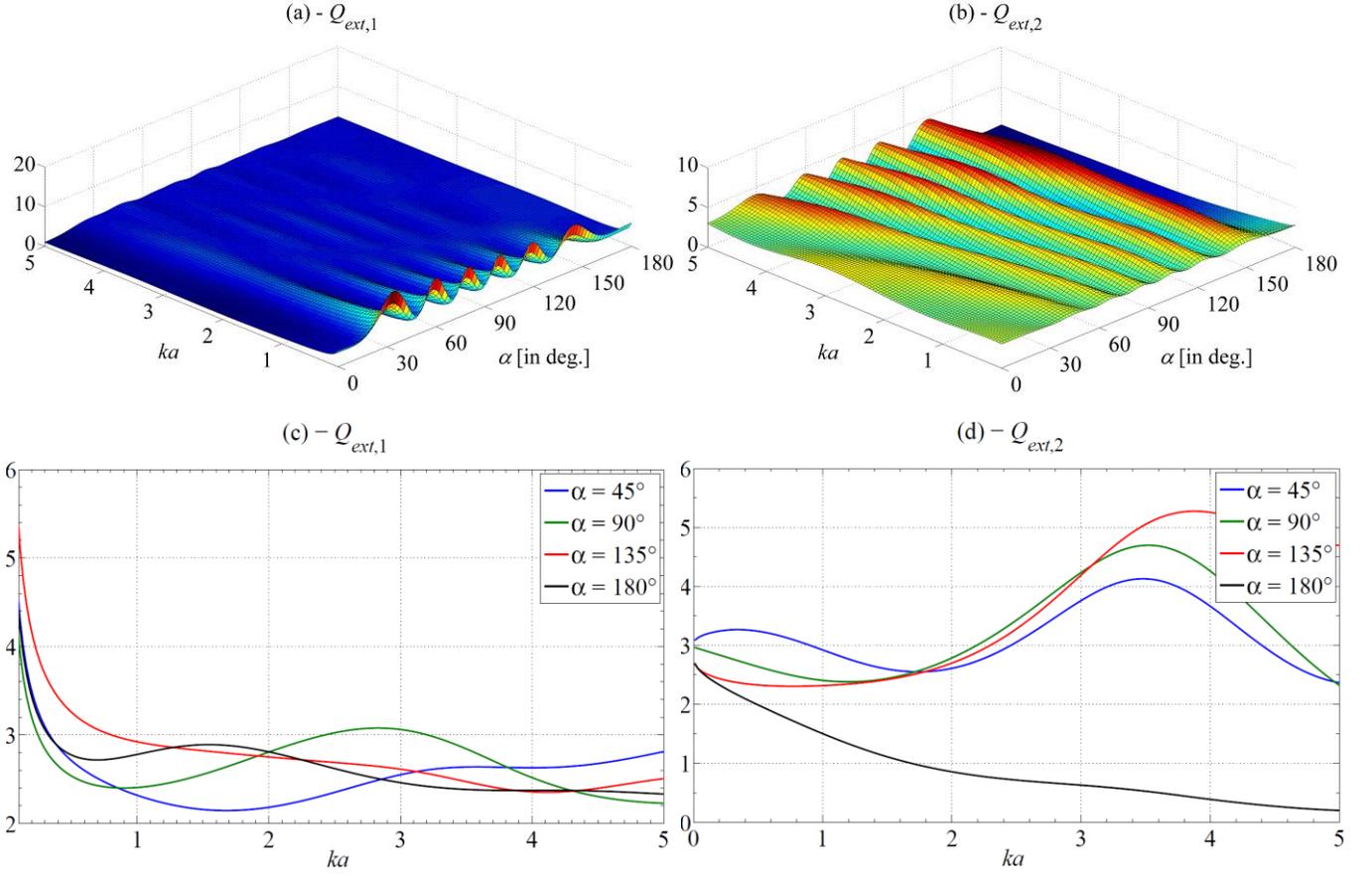

**Fig. 8.** The same as in Fig. 5, but $kb = 1$ and $kd = 20$.

Practically, in an experimental setting, *extrinsic* cross-sections are measured in the far-field [22, 23]. The extrinsic cross-sections are different from the local/intrinsic cross-sections, as shown in the acoustical context [27, 28, 37]. Extending the scope of the previously analysis for the extrinsic acoustical cross-sections [27], the EM *extrinsic* cross-sections (denoted by the superscript *e*) are obtained similarly as,

$$\sigma_{sca}^{e} = \frac{4}{k} \sum_{n=-\infty}^{\infty} \left[ |C_{n,1}|^2 + |C_{n,2}|^2 + 2\Re\left\{ C_{n,1}^{*} \sum_{m=-\infty}^{\infty} C_{m,2} J_{n-m}(kd) \right\} \right], \quad (27)$$

$$\sigma_{abs}^{e} = -\frac{4}{k} \Re\left\{ \sum_{n=-\infty}^{\infty} \left[ i^n e^{in\alpha} \left( C_{n,1} + \sum_{m=-\infty}^{\infty} C_{m,2} J_{n-m}(kd) \right) \right] \right\} - \sigma_{sca}^{e}, \quad (28)$$

$$\sigma_{ext}^{e} = -\frac{4}{k} \Re\left\{ \sum_{n=-\infty}^{\infty} \left[ i^n e^{in\alpha} \left( C_{n,1} + \sum_{m=-\infty}^{\infty} C_{m,2} J_{n-m}(kd) \right) \right] \right\}. \quad (29)$$

As noted, for example, from Eq.(27), the extrinsic scattering cross-section depends on the coefficients $C_{n,1}$ and $C_{n,2}$, which are *coupled*, i.e., they are a function of both objects simultaneously as shown in Eqs. (12) and (13). Nonetheless, there is an additional cross-scattering term described by the third factor in Eq.(27) with the dependence of the series on the cylindrical Bessel function of the first kind with argument $kd$. One aspect in which the computations for the local/intrinsic cross-sections (or efficiencies) presented in this analysis can be useful concerns the quantitative characterization of the additional cross-scattering term by separating the measured extrinsic cross-section from the intrinsic scattering cross-sections (for each object) computed *a priori*. Thus, in conjunction with the experimental data, the intrinsic cross-sections computed numerically *a priori* can be useful in the full characterization of a multiple scattering system of many particles.